\documentclass[a4paper, 10pt, 5p, preprint, twocolumn]{elsarticle}
\usepackage{amsmath,graphicx,color}
\usepackage[normalem]{ulem}
\newcommand{\trento}{T$\mathrel{\protect\raisebox{-2.1pt}{R}}$ENTo}

\begin{document}

\title{The mean transverse momentum of ultracentral heavy-ion collisions: \\  A new probe of hydrodynamics}

\author[1,2]{Fernando G. Gardim\corref{cor1}}
\author[2]{Giuliano Giacalone}
\author[2]{Jean-Yves Ollitrault}
\address[1]{Instituto de Ci\^encia e Tecnologia, Universidade Federal de Alfenas, 37715-400 Po\c cos de Caldas, MG, Brazil}
\address[2]{Universit\'e Paris Saclay, CNRS, CEA, Institut de physique th\'eorique, 91191 Gif-sur-Yvette, France}

\cortext[cor1]{Corresponding author}

\begin{keyword}
This article is registered under preprint number: /nucl-th/1909.11609'.
\end{keyword}

\begin{abstract}
 We predict that the mean transverse momentum of charged hadrons $\langle p_t\rangle$ rises as a function of the charged-particle multiplicity in ultracentral nucleus-nucleus collisions. We explain that this phenomenon has a simple physical origin and represents an unambiguous prediction of the hydrodynamic framework of heavy-ion collisions. We argue that the relative increase of $\langle p_t \rangle$ is proportional to the speed of sound squared $c_s^2$ of the quark-gluon plasma. Based on the value of $c_s^2$ from lattice QCD, we expect $\langle p_t\rangle$ to increase by approximately $18$~MeV between 1\% and 0.001\% centrality in Pb+Pb collisions at $\sqrt{s_{\rm NN}}=5.02$~TeV
\end{abstract}

\maketitle

\section{Introduction}
\label{s:introduction}

We predict a new phenomenon to be observed in experimental data on heavy-ion collisions: a rise of the mean transverse momentum of charged hadrons, $\langle p_t \rangle$, in ultracentral collisions.  The idea is that in the 0.1\% most central collisions the quark-gluon plasma has always the same volume, while the charged-particle multiplicity, $N_{\rm ch}$, can vary significantly, by as much as 10\%. The total entropy in the quark-gluon plasma is proportional to the multiplicity, therefore, at constant volume, the entropy {\it density\/}, $s$, is itself proportional to the multiplicity, and also varies by a few percent.  
As a consequence, in ultracentral collisions the temperature increases as a function of the multiplicity, which in turn implies a rise of the mean transverse momentum of charged hadrons~\cite{VanHove:1982vk}, $\langle p_t \rangle$, observed in the final state, due to tight correlation with the temperature~\cite{Gardim:2019xjs}. 

\begin{figure*}
    \centering
    \includegraphics[width=.85\linewidth]{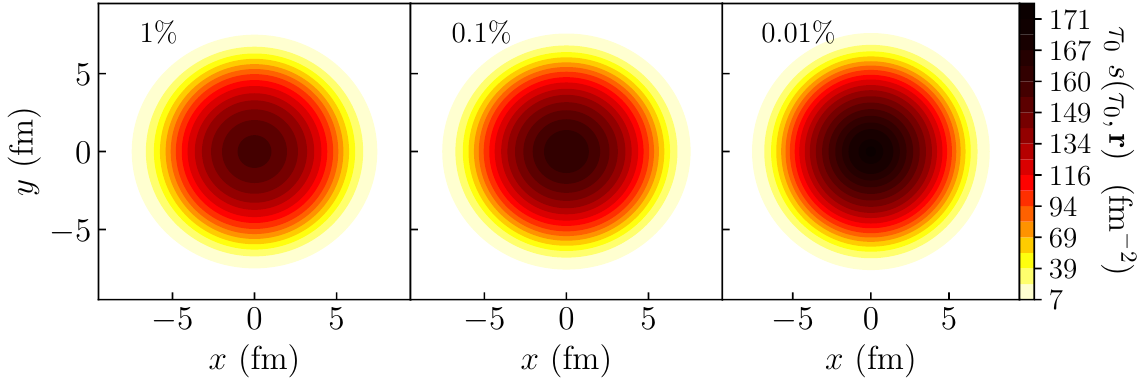}
    \caption{Entropy density per unit transverse area and unit rapidity, as a function of transverse coordinates, for three fixed centralities:  1\%, 0.1\% and 0.01\%. These profiles are constructed as follows: We first evaluate the rms radius of the entropy density profile and the mean impact parameter for each centrality using a high-statistics \trento{} calculation (see Sec.~\ref{s:quantitative}). We then construct a smooth profile corresponding to this mean impact parameter, where the entropy density is proportional to $\sqrt{T_AT_B}$, evaluated in an optical Glauber calculation, and we rescale the total entropy and the radius so that they match the full numerical calculation. In this way, we obtain profiles of entropy density which are essentially equivalent to the actual average profiles of entropy density that would be returned by a full numerical calculation at those centralities. The radius varies by roughly 1\% from the left to the right panel, while the entropy density varies by $\sim 13\%$.}
    \label{fig:fig0}
\end{figure*}

To illustrate the physical picture, we show in Fig.~\ref{fig:fig0} three entropy density profiles in the transverse plane, representing the average entropy density of Pb+Pb collisions at, respectively, 1\%, 0.1\% and 0.01\% centrality. 
One finds that these profiles have the same radius, but the entropy density (and so the temperature) increases as the collision becomes more central. Recent experimental analyses~\cite{Acharya:2018eaq} seem to contradict this prediction: $\langle p_t\rangle$ varies by less than $0.2\%$ in the 0-20\% centrality range in Pb+Pb collisions at $\sqrt{s_{\rm NN}}=5.02$~TeV. However, these analyses use wide centrality bins, while, as we shall see, the rise is only expected in ultracentral events. The observed flatness of $\langle p_t\rangle$ implies that even a modest rise in the ultracentral range~\cite{Luzum:2012wu,CMS:2013bza,Aaboud:2019sma} will be easy to identify.

This new phenomenon is expected to be driven by the speed of sound  $c_s$, which thus can be determined experimentally in ultrarelativistic heavy-ion collisions. The speed of sound is the velocity at which a compression wave travels in a fluid. Its magnitude is determined by the change in pressure as one increases the density. In a relativistic fluid, it is given by~\cite{Ollitrault:2008zz}

\begin{equation}
  \label{defcs}
c_s^2=\frac{dP}{d\epsilon}=\frac{d\ln T}{d\ln s},
\end{equation}
where $P$, $\epsilon$, $T$, $s$ denote, respectively, the pressure, energy density, temperature, and entropy density. Assuming that $\langle p_t\rangle$ is proportional to the temperature~\cite{VanHove:1982vk,Gardim:2019xjs}, this rise allows one to determine the speed of sound directly as a function of experimental quantities using
\begin{equation}
\label{csultraultra}
  c_s^2=\frac{d\ln\langle p_t\rangle}{d\ln N_{\rm ch}}.
\end{equation}
This analysis requires to bin events in $N_{\rm ch}$ or, equivalently, to determine the centrality using $N_{\rm ch}$~\cite{Aaboud:2019sma}. 
We use $N_{\rm ch}$ as a measure of the entropy and $\langle p_t\rangle$ as a measure of the temperature.
Consistency then requires that both should be measured in the same rapidity window, at variance with current analyses where the centrality is typically determined in a separate rapidity window~\cite{Abelev:2013qoq}.

The physics of ultracentral collisions is not yet fully understood. 
The ratio of elliptic flow to triangular flow is smaller than predicted by models~\cite{Carzon:2020xwp}, and elliptic flow fluctuations display an irregular behaviour, with the fourth cumulant changing sign as a function of centrality~\cite{Aaboud:2019sma}.
The theoretical description of anisotropic flow, however, involves the detailed modeling of the initial stages of the collision. 
By contrast, the mean transverse momentum discussed in this Letter is determined by conservation laws~\cite{Gardim:2019xjs}, and model details are to a large extent irrelevant.

We make a quantitative prediction for the increase of $\langle p_t\rangle$  in ultracentral Pb+Pb collisions at $\sqrt{s_{\rm NN}}=5.02$~TeV, which is parameter-free and does not rely on any specific model. 
We use a specific model as an illustration only:
In Sec.~\ref{s:quantitative}, we evaluate quantitatively the centrality dependence of the entropy density, $s$, using the \trento{}  model of initial conditions~\cite{Moreland:2014oya}.
We then identify, in Sec.~\ref{s:analytic}, features which are general and do not rely on this particular model. 
We propose a refinement of Eq.~(\ref{csultraultra}) that solely involves information inferred from the measured distribution of $N_{\rm ch}$~\cite{Aaboud:2019sma}. 
The resulting prediction is presented in Sec.~\ref{s:prediction}. 

\begin{figure}[t]
    \centering
    \includegraphics[width=\linewidth]{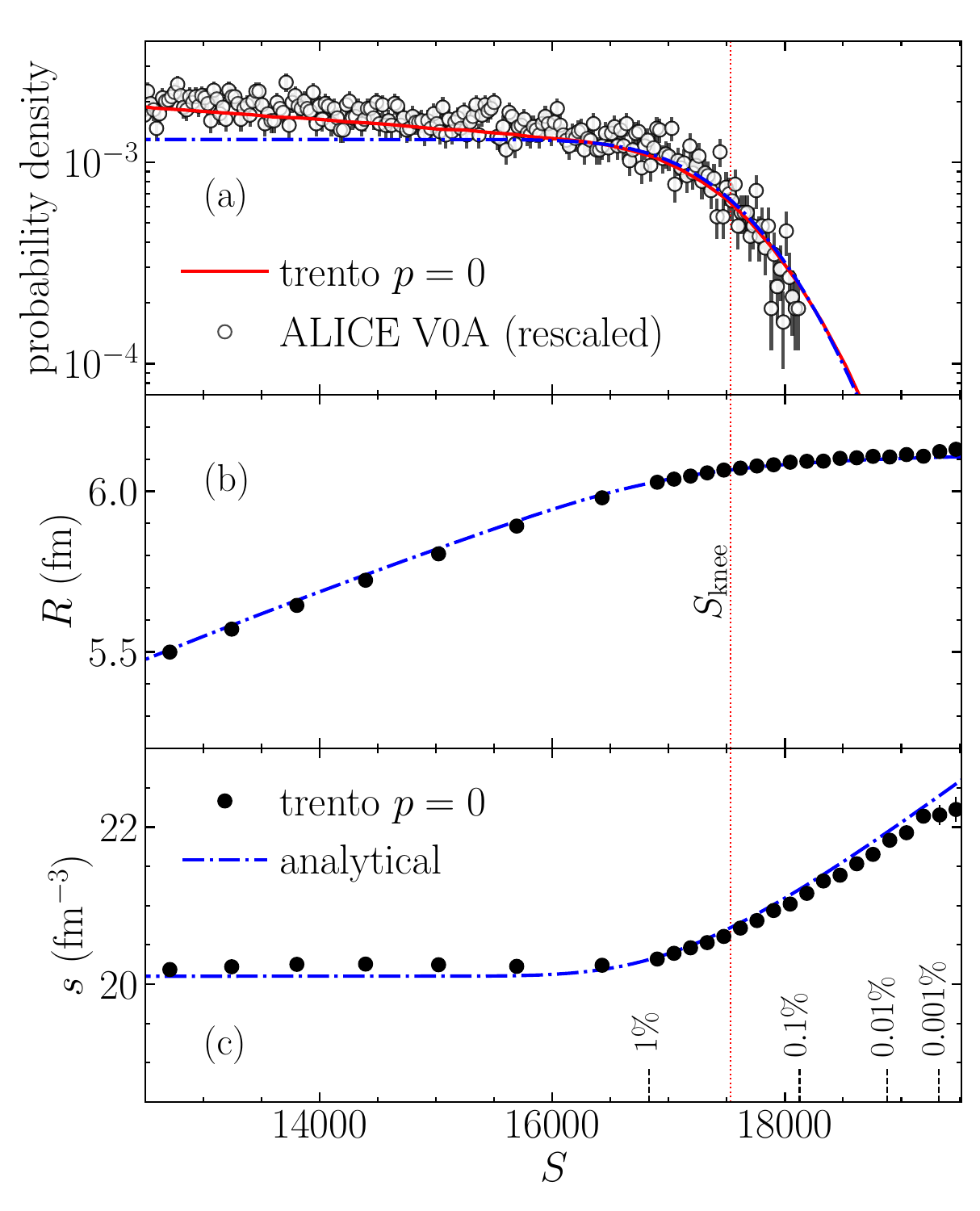}
    \caption{Results from the \trento{} model of initial conditions~\cite{Moreland:2014oya}, with $p=0$ and $k=2.0$.
      20 million Pb+Pb collisions at $\sqrt{s_{\rm NN}}=5.02$~TeV were generated. 
      Only 10\% of these events, corresponding to the largest values of the total entropy per unit rapidity $S$ (0-10\% centrality), are used. 
      (a) Full line: Probability distribution of $S$ in the \trento{} calculation. Open symbols: Probability distribution of the V0 amplitude, used by the ALICE Collaboration to determine the centrality~\cite{Adam:2015ptt}, rescaled by a factor $0.51$. 
      (b) Values of the initial radius $R$, given by Eq.~(\ref{defR}). (c) Entropy density, $s\propto S/R^3$.
The proportionality constant has been chosen so that the entropy density left of the knee is $20.1$~fm$^{-3}$~\cite{Gardim:2019xjs}.      
Symbols in panels (b) and (c) are the results of the \trento{}  simulations averaged over events. 
Dot-dashed lines in panels (a), (b) and (c) are one-parameter fits using Eqs.~(\ref{modela}), (\ref{modelb}) and (\ref{modelc}). 
Vertical lines spot specific values of the centrality percentile, and the position of the knee. 
    }
    \label{fig:fig1}
\end{figure}

\section{Quantitative analysis}
\label{s:quantitative}

We use the \trento{} Monte Carlo generator of initial conditions~\cite{Moreland:2014oya} with the $p=0$ prescription (corresponding to an entropy density proportional to $\sqrt{T_AT_B}$, where $T_A$ and $T_B$ are the thickness functions of incoming nuclei~\cite{Miller:2007ri}), which has been employed successfully in phenomenological applications~\cite{Giacalone:2017dud}.

As we shall argue in Sec.~\ref{s:analytic}, the details of the model are irrelevant. 
It is however essential that the model has the right multiplicity fluctuations, since the predicted increase of $\langle p_t\rangle$ is due to these fluctuations. 
Therefore, we tune the fluctuation parameter of \trento{}, $k$, in such a way that the distribution of the total entropy per unit rapidity, $S$, coincides, up to a global multiplicative constant, with the distribution of the multiplicity (V0 amplitude) used by the ALICE Collaboration to define the centrality in Pb+Pb collisions at $\sqrt{s_{\rm NN}}=5.02$~TeV~\cite{Adam:2015ptt}. 
The same choice of parameters also reproduces the distribution of $N_{\rm ch}$ measured by the ATLAS Collaboration~\cite{Aaboud:2019sma}.
We rescale the entropy given by the \trento{} model so as to match the value of the entropy per unit rapidity inferred from the measured charged multiplicity~\cite{Adam:2015ptt}, using $S=6.7N_{\rm ch}$~\cite{Hanus:2019fnc}.\footnote{We correct for the Jacobian transformation between pseudorapidity $\eta$ and rapidity $y$ using $dN_{\rm ch}/dy\simeq  1.15 dN_{\rm ch}/d\eta$~\cite{Hanus:2019fnc}.}

The distribution of $S$ in the \trento{} model is displayed as a solid line in Fig.~\ref{fig:fig1} (a).
Symbols indicate the distribution of the quantity used by the ALICE Collaboration to determine the centrality~\cite{Abelev:2013qoq,Adam:2015ptt}. This quantity is proportional to the total entropy, and has been rescaled by a global factor so as to match the distribution of $S$ in \trento{}. 
The histogram displays two regimes, left and right of the {\it knee}, which is indicated by a vertical line, and which will be defined below in Eq.~(\ref{defknee}).
Left of the knee, the distribution decreases slowly. The variation of $S$ in this region is driven by the variation of impact parameter.
Right of the knee, the distribution decreases steeply.
In this region, the variation of $S$ is driven by initial-state fluctuations. 

Next, we calculate the transverse radius, $R$, which is defined in a given event by 
\begin{equation}
  \label{defR}
  R^2\equiv 2\left(\langle {\bf r}^2\rangle-\langle {\bf r}\rangle^2\right),
\end{equation}
where ${\bf r}=(x,y)$ is the transverse coordinate, and angular brackets denote an average value taken with the initial entropy density as a weight.\footnote{The factor 2 in Eq.~(\ref{defR}) ensures that the right-hand side is equal to $R^2$ if the entropy density is uniform in a circle of radius $R$.}
Figure~\ref{fig:fig1} (b) displays the value of $R$, averaged over events, as a function of $S$.
It increases and then roughly saturates to a constant value when $S\approx S_{\rm knee}$. This confirms the intuitive idea that the events beyond the knee share the same geometry. 

\begin{figure}[t]
    \centering
    \includegraphics[width=\linewidth]{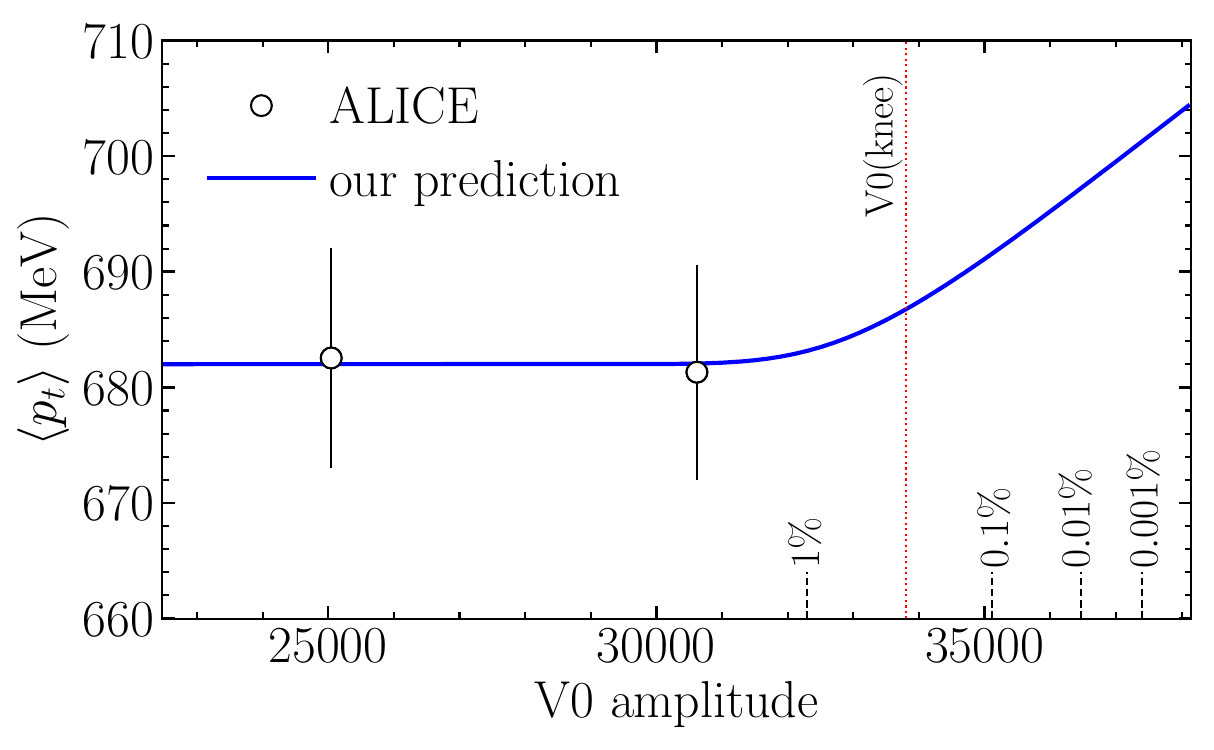}
    \caption{Line: our prediction for the variation of $\langle p_t\rangle$ with the V0 amplitude in Pb+Pb collisions at $\sqrt{s_{\rm NN}}=5.02$~TeV.
      Symbols are data from the ALICE collaboration~\cite{Acharya:2018eaq}. 
    }
    \label{fig:fig2}
\end{figure}

Finally, we calculate the entropy density, $s$, which is proportional to $S/R^3$ for dimensional reasons.
Figure~\ref{fig:fig1} (c) displays its value averaged over events.
Left of the knee, the entropy density is almost constant, which in turn implies that the temperature and the mean transverse momentum $\langle p_t\rangle$ are also constant, in agreement with experimental data (see below Fig.~\ref{fig:fig2}). 
The essential observation of this paper is that, right of the knee, the entropy density starts rising because the volume becomes constant, so that $s$ becomes proportional to $S$. 

\section{Analytic model}
\label{s:analytic}

We now derive a simple parametrization which captures the trends observed in Fig.~\ref{fig:fig1}.
We assume that event-to-event fluctuations of $S$ at a fixed impact parameter, $b$, are Gaussian~\cite{Das:2017ned}:
\begin{equation}
  \label{gaussian}
  P(S|b)=\frac{1}{\sigma\sqrt{2\pi}}\exp\left(-\frac{(S-\bar S(b))^2}{2\sigma^2}\right),
\end{equation}
where $\bar S(b)$ is the mean value, which decreases with increasing $b$, and $\sigma$ is the width, whose dependence on $b$ can be neglected since we focus on events in a narrow bin of centrality. 
The knee of the histogram of $S$ is defined as the mean value of the entropy at $b=0$~\cite{Das:2017ned}:
\begin{equation}
  \label{defknee}
  S_{\rm knee}\equiv \bar S(0). 
\end{equation}
We first derive the distribution of $S$ by integrating over impact parameter. 
We perform the change of variable $b \to \bar S(b)$, so that Eq.~(\ref{gaussian}) becomes
\begin{equation}
  \label{gaussian2}
  P(S|\bar S)=\frac{1}{\sigma\sqrt{2\pi}}\exp\left(-\frac{(S-\bar S)^2}{2\sigma^2}\right).
\end{equation}
We then integrate over $\bar S$:
\begin{eqnarray}
  \label{modela}
  P(S)&=&\int_0^{S_{\rm knee}} P(S|\bar S)P(\bar S)d\bar S\cr
&\propto&\int_0^{S_{\rm knee}} P(S|\bar S)d\bar S\cr
&\propto& {\rm erfc}\left(\frac{S-S_{\rm knee}}{\sigma\sqrt{2}}\right),
\end{eqnarray}
where we have assumed for simplicity that the probability distribution of $\bar S$, $P(\bar S)$, is constant. 
 The distribution of $S$ obtained in this model is displayed as a dot-dashed line in Fig.~\ref{fig:fig1} (a).
The parameters $S_{\rm knee}$ and $\sigma$ have been obtained within the \trento{} model by computing the mean and standard deviation of the distribution of $S$ at $b=0$.
The values are $S_{\rm knee}=17554$ and $\sigma=674$. 
The overall proportionality constant in Eq.~(\ref{modela}) is adjusted by hand. 
This simple model captures the trends observed in the \trento{} simulation up to 10\% centrality.

Next, we assume that the initial radius, $R$, only depends on impact parameter, or equivalently, on $\bar S$. 
In order to determine $R$ for fixed $S$, we
first determine the distribution of $\bar S$ for fixed $S$ using 
Bayes' theorem:
\begin{equation}
  \label{bayes}
P(\bar S|S)=\frac{P(S|\bar S)P(\bar S)}{P(S)}. 
\end{equation}
The average value of $\bar S$ for fixed $S$, denoted by $\langle \bar S|S\rangle$, is obtained by inserting Eq.~(\ref{gaussian2}) into Eq.~(\ref{bayes}) and integrating over $\bar S$. Assuming again that $P(\bar S)$ is constant, we obtain:
  \begin{equation}
    \label{modeli}
    \langle \bar S|S\rangle=
    S-\sigma\sqrt{\frac{2}{\pi}}
    \frac{\exp\left(-\frac{(S-S_{\rm knee})^2}{2\sigma^2}\right)}{{\rm erfc}\left(\frac{S-S_{\rm knee}}{\sqrt{2}\sigma}\right)}. 
    \end{equation}
  For $S<S_{\rm knee}$, the second term in the right-hand side is negligible, and
  $\langle \bar S|S\rangle\simeq S$, i.e., fluctuations are averaged out~\cite{Broniowski:2001ei}. 
  Right of the knee, $\bar S$ saturates to its maximum value: $\langle \bar S|S\rangle\simeq S_{\rm knee}$. 

  The observation that the entropy density is constant left of the knee in the \trento{} calculation suggests that the volume is proportional to $\bar S$.
  Under this assumption, the radius $R$ is given by
  \begin{equation}
    \label{modelb}
    R=R_0\left(\frac{\langle \bar S|S\rangle}{S_{\rm knee}}\right)^{1/3}
  \end{equation}
  while the entropy density is given by
  \begin{equation}
    \label{modelc}
    s=s_0\frac{S}{\langle \bar S|S\rangle}.
       \end{equation}
  In these equations, $R_0$ and $s_0$ are correspond to the value of $R$ right of the knee, and the value of $s$ left of the knee, respectively.
  Dot-dashed lines in Fig.~\ref{fig:fig1} (b) and  Fig.~\ref{fig:fig1} (c) are fits to the full \trento{} simulation using Eqs.~(\ref{modelb}) and (\ref{modelc}).
  There is only one fit parameter for each curve, $R_0$ for Fig.~\ref{fig:fig1} (b) and $s_0$ for Fig.~\ref{fig:fig1} (c), and the quality of the fit is excellent.
  This implies that the centrality dependence of the entropy density is captured by Eq.~(\ref{modelc}).
Now, the model of initial conditions enters this equation only through the global proportionality constant, $s_0$. 
We conclude that the validity of Eq.~(\ref{modelc}) is more general than the particular model used to test it, and we expect that it would be valid also in other models commonly used for the description of initial state fluctuations, such as IP Glasma\cite{Schenke:2012wb} or EbyE EKRT~\cite{Niemi:2015qia}.
  
\section{Quantitative predictions for Pb+Pb collisions at $\sqrt{s_{\rm NN}}=5.02$~TeV}
\label{s:prediction}

  We now make quantitative predictions using Eq.~(\ref{modelc}).
  The interest is that all the parameters can be determined from data.
More specifically, one replaces $S$ with the charged-particle multiplicity, $N_{\rm ch}$, so that the quantities $S_{\rm knee}$ and $\sigma$ can be determined from the distribution of $N_{\rm ch}$.
This can be done either using the simple Bayesian procedure of Ref.~\cite{Das:2017ned},
or by fitting a model (such as the Glauber model) to the experimental histogram and computing $S_{\rm knee}$ and $\sigma$ in this model.
Here we apply the fitting procedure of Ref.~\cite{Das:2017ned} to ALICE data, using the V0 amplitude as a proxy for the charged multiplicity~\cite{Abelev:2013qoq} and using the same data shown in Fig.~\ref{fig:fig1} (a).
We obtain $S_{\rm knee}=33800$ (denoted by V0(\rm knee) in Fig.~\ref{fig:fig2}) and $\sigma=1140$.

  We next assume that the mean transverse momentum is proportional to the temperature, which is itself proportional to $s^{c_s^2}$ if the temperature range is narrow enough that one can neglect the variation of $c_s^2$.
Using Eq.~(\ref{modelc}), we obtain the prediction:
\begin{equation}
  \label{prediction}
\langle p_t\rangle=p_{t0}\left(\frac{S}{\langle \bar S|S\rangle}\right)^{c_s^2}, 
\end{equation}
where $p_{t0}$ is the value of $\langle p_t\rangle$ left of the knee, and $\langle \bar S|S\rangle$ is given by Eq.~(\ref{modeli}). 
We use the value $p_{t0}=682$~MeV measured by ALICE in the 0-5\% centrality range~\cite{Acharya:2018eaq}, and we take the value of $c_s^2$ from lattice QCD~\cite{Borsanyi:2013bia} calculations.
The velocity of sound depends on the temperature, but $\langle p_t \rangle$ in central Pb+Pb collisions at $\sqrt{s_{\rm NN}}=5.02$~TeV probes the equation of state around an effective temperature $T=222$~MeV~\cite{Gardim:2019xjs}, at which lattice QCD gives $c_s^2=0.252$. 
This yields the prediction displayed in Fig.~\ref{fig:fig2}.
We predict that $\langle p_t\rangle$ increases by $8.4$~MeV between $1\%$ and $0.1\%$ centrality, by $5.6$~MeV between $0.1\%$ and $0.01\%$, and by $4.1$~MeV between $0.01\%$ and $0.001\%$. 

Note that our prediction does not rely on any specific model of the collision.
The sole physics assumption is that the quantity used to measured the centrality (e.g. the V0 amplitude for the ALICE experiment) is proportional to the entropy of the system on an event-by-event basis.
This is only approximately true for two reasons: 
First, in the case of ALICE, the V0 amplitude and $\langle p_t\rangle$ are measured intwo different rapidity windows, and entropy fluctuations may depend on rapidity. 
Second, the observed fluctuations of multiplicity get a small contribution from trivial statistical (Poisson) fluctuations, which do not contribute to the rise of $\langle p_t\rangle$. 
In the case of ATLAS data~\cite{Aaboud:2019sma} on the distribution of $N_{\rm ch}$, the width of Poisson fluctuations is smaller by a factor $2.5$ than the total width.
Assuming that statistical and dynamical fluctuations add up in quadrature, this implies that the width of dynamical fluctuations is 90\% of the total width.
Thus, one expects a 10\% reduction of the rise of $\langle p_t\rangle$ due to trivial statistical fluctuations. 

Equation~(\ref{prediction}) reduces to Eq.~(\ref{csultraultra}) for the most central events, where $\langle\bar S|S\rangle\simeq S_{\rm knee}$, if one replaces $S$ with $N_{\rm ch}$. 
Its advantage over Eq.~(\ref{csultraultra}) is that it can be used all the way up to 10\% centrality. 
Experimentally, $c_s$ can be measured by fitting Eq.~(\ref{prediction}) to data, using $p_{t0}$ and $c_s$ as fit parameters.
Such an analysis would complement the extraction of $c_s$ from the variation of $\langle p_t\rangle$ with $\sqrt{s_{\rm NN}}$~\cite{Gardim:2019xjs}.
But more importantly, the predicted rise of $\langle p_t\rangle$ in ultra-central collisions provides a nontrivial test of the hydrodynamic behavior of nucleus-nucleus collisions which does not involve anisotropic flow~\cite{Heinz:2013th}. 

\section*{Acknowledgments}
F.G.G. was supported by Conselho Nacional de Desenvolvimento Cient\'{\i}fico  e  Tecnol\'ogico  (CNPq grant 205369/2018-9 and 312932/2018-9). 
F.G.G. acknowledges support from project INCT-FNA Proc.~No.~464898/2014-5.

\end{document}